\pdfoutput=1
\documentclass[a4paper,fleqn,usenatbib]{mnras}
\usepackage[T1]{fontenc}
\usepackage{ae,aecompl}
\usepackage{graphicx}	
\usepackage{amsmath}	
\usepackage{amssymb}
\usepackage{xcolor}
\definecolor{notecolor}{rgb}{0.8,0,0}

\def\lya{Ly$\alpha$~}
\def\c2{[C~\textsc{ii}]}
\def\h1{H~\textsc{i}}
\def\kunit{$h/$cMpc}

\title[High-redshift Intensity Mapping]{Predictions and sensitivity
  forecasts for reionization-era [C~{\sc ii}] line intensity mapping}

\author[Dumitru et al.]
       {{Sebastian Dumitru$^{1,2,3}$\thanks{Email: sdumitru@sas.upenn.edu},
           Girish Kulkarni$^{1,2,4}$\thanks{Email: kulkarni@theory.tifr.res.in},
           Guilaine Lagache$^{5}$}
         \newauthor{and Martin G. Haehnelt$^{1,2}$}\\
         $^1$Institute of Astronomy and Kavli Institute of Cosmology,
         University of Cambridge, Madingley Road, Cambridge CB3 0HA, UK\\
         $^2$Kavli Institute of Cosmology, University of Cambridge,
         Madingley Road, Cambridge CB3 0HA, UK \\
         $^3$Department of Physics and Astronomy, University of
         Pennsylvania, Philadelphia PA 19104, USA\\
         $^4$Department of Theoretical Physics, Tata Institute of
         Fundamental Research, Homi Bhabha Road, Mumbai 400005,
         India\\
         $^5$Aix Marseille Univ, CNRS, LAM, Laboratoire
         d'Astrophysique de Marseille, Marseille, France
} 

\date{Accepted ---. Received ---; in original form ---}

\pubyear{2017}

\begin{document}
\label{firstpage}
\pagerange{\pageref{firstpage}--\pageref{lastpage}}
\maketitle

\begin{abstract}
Observations of the high-redshift Universe using the 21~cm line of
neutral hydrogen and complimentary emission lines from the first
galaxies promise to open a new door for our understanding of the epoch
of reionization.  We present predictions for the \c2\ 158~$\mu$m line
and \h1 21~cm emission from redshifts $z=6$--$9$ using
high-dynamic-range cosmological simulations combined with
semi-analytical models.  We find that the CONCERTO experiment should
be able to marginally detect the large scale power spectrum of \c2\
emission to redshifts of up to $z=8$ (signal-to-noise ratio $\sim
1$ at $k< 0.1$~\kunit\ with 1500~hr of integration).  A Stage~II
experiment similar to CCAT-p should be able to detect \c2\ from even
higher redshifts to high significance for similar integration times
(signal-to-noise ratio of $\sim 50$ at $k= 0.2$~\kunit\ at $z=6$).  We
study the possibility of combining such future
\c2\ measurements with 21~cm measurements using LOFAR and SKA to
measure the \c2-21cm cross power spectra, and find that a Stage~II
experiment should be able to measure the cross-power spectrum for
$k\lesssim 1$~\kunit\ to signal-to-noise ratio of better than 10.  We
discuss the capability of such measurements to constrain astrophysical
parameters relevant to reionization and show that a measurement of the
\c2-21cm cross power spectrum helps break the degeneracy between the
mass and brightness of ionizing sources.
\end{abstract}


\section{Introduction}

Atomic and molecular emission lines with wavelength redward of
hydrogen Lyman-$\alpha$ have the desirable property of remaining
visible deep into the epoch of hydrogen reionization (redshift
$z=6$--$10$), where the \lya line is difficult to observe due to
saturated absorption.  These emission lines, which depend on the cold
gas content, the ionising radiation field, or the metallicity,
uniquely probe the formation of the very first stars and
galaxies. They should be a good tracer of the cosmic density
structure.

Intensity mapping of such emission lines (e.g., O~\textsc{i},
O~\textsc{iii}, C~\textsc{ii}, CO, H~\textsc{i}, H$_2$) is an
attractive tool to study the high-redshift Universe
\citep{1999ApJ...512..547S, 2010JCAP...11..016V, 2011ApJ...730L..30C,
  2011ApJ...728L..46G, 2011ApJ...741...70L, 2012ApJ...745...49G,
  2013ApJ...768..130G, 2015ApJ...806..209S, 2015MNRAS.450.3829Y,
  2016ApJ...833..153S, 2017MNRAS.464.1948F}.  By measuring large-scale
variations in line emission from many individual unresolved galaxies,
intensity mapping provides a statistical measurement that encodes
cosmological and astrophysical information.  This capacity of
intensity mapping experiments is particularly important at redshifts
corresponding to the epoch of reionization, which is a key period in
the history of the Universe, when the earliest galaxies and quasars
form and ionize the surrounding neutral hydrogen.  Constraints from
the evolution in the Ly$\alpha$ opacity of the intergalactic medium
(IGM; e.g., \citealt{2006AJ....132..117F, 2017ApJ...844...85O}) and
the temperature and polarization anisotropy in the cosmic microwave
background (CMB; \citealt{2016A&A...596A.108P}) suggest that
reionization occurs at redshifts $z\sim 6$--$15$.  However, the nature
of the sources of reionization remains uncertain.  Measurements of the
escape fraction of Lyman-continuum photons necessary for reionization
from high-redshift galaxies are still elusive.  Although galaxies down
to rest-frame UV magnitudes of $M_\mathrm{UV}=-12.5$ ($L\sim
10^{-3}L^*$) at redshift $z=6$ \citep{2017ApJ...835..113L} and
redshifts as high as $z=11.1$ \citep{2016ApJ...819..129O} have been
observed, the escape fraction of Lyman-continuum photons has been
measured in only a handful of bright ($L>0.5L^*$) and low-redshift ($z
< 4$) galaxies.  In these galaxies, the escape fraction is typically
found to be 2--20\% \citep{2010ApJ...725.1011V, 2011ApJ...736...41B,
  2015ApJ...804...17S, 2015ApJ...810..107M, 2016A&A...585A..48G,
  2017MNRAS.468..389J, 2017MNRAS.465..316M} but reionization requires
escape fractions of about 20\% in galaxies down to $M_\mathrm{UV}=-13$
\citep{2016PASA...33...37F, 2015ApJ...802L..19R, 2016MNRAS.457.4051K}.
There is tentative evidence for a dominant contribution to
reionization from quasars from the suggestion of a rather steep faint
end of the QSO luminosity function at high redshift by
\citet{2015A&A...578A..83G}, and large \lya opacity fluctuations at
very large scales in QSO absorption spectra
\citep{2015MNRAS.447.3402B,2015MNRAS.453.2943C, 2016MNRAS.460.1328D}.
But it may be difficult to reconcile this with measurements of the
He~\textsc{ii} \lya opacity and measurements of the IGM temperature at
$z\sim 3$ \citep{2018arXiv180104931P, 2017MNRAS.468.4691D,
  2015ApJ...813L...8M}, and also with measurements of the incidence
rate of metal-line systems \citep{2016MNRAS.459.2299F}.

Intensity mapping of atomic and molecular lines emission from galaxies
in the epoch of reionization has the potential to unambiguously reveal
the properties of the sources of reionization.  The radiative transfer
of emission in these lines in galaxies is very different from that of
the Lyman-continuum emission.  As a result, intensity mapping yields a
view of high-redshift galaxies that is unbiased by their
Lyman-continuum escape fraction.  Cross-correlating this measurement
with a measurement of the ionization state of the large-scale IGM,
such as of the 21~cm emission or absorption from the IGM, can then
result in constraints on reionizing sources.

Several experiments are currently in deployment to measure the
large-scale clustering in the 21~cm signal from the IGM during the
epoch of reionization, such as Murchison Widefield Array (MWA;
\citealt{2013PASA...30...31B, 2013PASA...30....7T}), Low Frequency
Array (LOFAR; \citealt{2013A&A...556A...2V};
\citealt{2014ApJ...782...66P}), Hydrogen Epoch of Reionization Array
(HERA; \citealt{2014ApJ...782...66P}, \citealt{2016arXiv160607473D}),
and Square Kilometre Array (SKA; astronomers.skatelescope.org).
However, 21~cm power spectrum observations alone are limited in their
capability of constraining reionization parameters.  This is due to
the degeneracy between the Lyman-continuum escape fraction (sometimes
also parameterised as the ionization efficiency) and the mass of
ionizing sources: a wide range in the host halo masses of ionizing
sources can produce very similar large-scale 21~cm power for a variety
of escape fraction values \citep{2015MNRAS.449.4246G}.
Cross-correlations with other line intensity maps can potentially
solve this problem by breaking the degeneracy. Our aim in this paper
is to investigate this possibility.

Various emission lines have been considered in the literature as
candidates for high-redshift intensity mapping, such as Ly$\alpha$
\citep{2013ApJ...763..132S, 2014ApJ...786..111P}, [O~\textsc{i}]
63.2~$\mu$m and 145.5~$\mu$m \citep{ 2011JCAP...08..010V,
  2016ApJ...833..153S}, CO(1--0) 2601~$\mu$m
\citep{2011ApJ...741...70L, 2011ApJ...728L..46G}, [N~\textsc{II}]
121.9~$\mu$m and 205.2~$\mu$m \citep{2016ApJ...833..153S} and \c2
157.6~$\mu$m \citep{2012ApJ...745...49G, 2015ApJ...806..209S,
  2016ApJ...833..153S}.  As these lines are a result of a reprocessing
of stellar emission by the interstellar medium
\citep{2001PhR...349..125B, 2013ARA&A..51..105C}, we generally expect
an anti-correlation on large scales between their signal and that of
the 21cm line, which originates in the neutral regions far away from
galaxies \citep{2011ApJ...741...70L}.  The intensity mapping technique
has been used at $z\sim 0.8$ using the 21~cm
line \citep{2010Natur.466..463C}, and at $z\sim 3$ using the \c2
\citep{2017arXiv170706172P} and CO \citep{2016ApJ...830...34K} lines.
Surveys suggested for future intensity mapping include CO Mapping
Pathfinder \citep{2016ApJ...817..169L} for CO at redshifts $z\sim
2$--$3$; TIME \citep{2014SPIE.9153E..1WC} and CONCERTO
\citep{2018arXiv180108054L, 2016ApJ...833..153S} for \c2\ at redshifts
$z=5$--$9$; HETDEX \citep{2008ASPC..399..115H} for \lya at
$z=1.9$--$3.5$; SPHEREx for Ly$\alpha$ at redshift $z\sim 6$--$8$ and
other lines at lower redshifts \citep{2014arXiv1412.4872D,
  2016arXiv160607039D}, and CDIM \citep{2016arXiv160205178C} for
H$\alpha$, O~III, and Ly$\alpha$ at $z=0.2$--$10$.

In this paper, we present predictions for \c2\ and 21\,cm brightness
power spectra and the \c2--21cm cross-power spectra from the epoch of
reionization ($z=6$--$10$) using a high-dynamic-range cosmological
hydrodynamical simulation from the Sherwood simulation suite
\citep{2017MNRAS.464..897B}.  We forecast the sensitivity to measure
these statistical quantities for the CONCERTO experiment
\citep{2018arXiv180108054L, 2016ApJ...833..153S} as well as a Stage~II
successor experiment beyond TIME and CONCERTO for \c2. For \h1, we use
the experimental setups of LOFAR and SKA.  Finally, we discuss the
feasibility of such experiments to constrain key parameters by
considering simple models of reionization.  The paper is organized as
follows. We first present our \c2 emission line model and 21\,cm line
maps in Sections~\ref{sec:model} and \ref{sec:HI_model}, and then
compute the cross-correlation between the \c2 and the 21~cm lines from
the epoch of reionization in Section~\ref{sec:cross-corr}.  We discuss
the observability of the \c2 and 21~cm power spectra and the \c2--21cm
cross power spectrum in Section~\ref{sec:im}. Finally, we illustrate
in Section~\ref{sec:forecasts} how the cross-correlation can be used
to probe the nature of ionizing sources, using in particular two
quantities: the minimum halo mass corresponding to a non-zero
Lyman-continuum photon escape fraction and the number of ionizing
photons produced by a halo.  We end by summarising our results in
Section~\ref{sec:conclusions}.  Our $\Lambda$CDM cosmological model
has $\Omega_\mathrm{b}=0.0482$, $\Omega_\mathrm{m}=0.308$,
$\Omega_\Lambda=0.692$, $h=0.678$, $n=0.961$, $\sigma_8=0.829$, and
$Y_\mathrm{He}=0.24$ \citep{2014A&A...571A..16P}.

\section{[C~\textsc{ii}] emission from high-redshift galaxies}
\label{sec:model}

We use a hydrodynamical cosmological simulation to model the \c2\ and
21~cm signal from the epoch of reionization.  This underlying
simulation is identical to that used in previous
work \citep{2016MNRAS.463.2583K, 2017MNRAS.469.4283K}, and is part of
the Sherwood simulation suite
(nottingham.ac.uk/astronomy/sherwood; \citealt{2017MNRAS.464..897B}).
It has been run using the energy- and entropy-conserving TreePM
smoothed particle hydrodynamical (SPH) code
\textsc{p-gadget-3} that is derived from the publicly available 
\textsc{gadget-2} code \citep{2001NewA....6...79S,
  2005MNRAS.364.1105S}.  We perform this simulation in a periodic,
  cubic volume that is 160 $h^{-1}$cMpc long.  A large dynamic range
  was achieved by using a softening length of $l_\mathrm{soft}=3.13$
  $h^{-1}$ckpc, and $2048^3$ dark matter and gas particles.  This
  results in a dark matter particle mass of $M_\mathrm{dm}=3.44\times
  10^7$ $h^{-1}$M$_\odot$ and gas particle mass of
  $M_\mathrm{gas}=6.38\times 10^6$ $h^{-1}$M$_\odot$.  Initial
  conditions were set up at redshift $z=99$ and evolved down to $z=4$.
  We saved simulation snapshots at 40~Myr intervals between $z=40$ and
  $z=4$.  In this paper, we use snapshots at $z=6.3, 7.1, 8.2$ and
  $9$.  In order to speed the simulation up, galaxy formation is
  simplified by using the {\tt QUICK\_LYALPHA} implementation
  in \textsc{p-gadget-3}.  This converts gas particles with
  temperature less than $10^5$~K and overdensity of more than a
  thousand times the mean baryon density to collisionless
  stars \citep{2004MNRAS.354..684V}.  Ionisation and thermal state of
  the gas is derived by solving for the ionization chemistry under the
  assumption of an equilibrium with the metagalactic UV background
  modelled according to \citet{2012ApJ...746..125H}.  The UV
  background of \citet{2012ApJ...746..125H} is slightly modified to
  result in IGM temperatures that agree with measurements by
\citet{2011MNRAS.410.1096B}.  This chemistry solver assumes radiative
  cooling via two-body processes such as collisional excitation of
H~\textsc{i}, He~\textsc{i}, and He~\textsc{ii}, collisional
ionization of H~\textsc{i}, He~\textsc{i}, and He~\textsc{ii},
recombination, and Bremsstrahlung
\citep{1996ApJS..105...19K}, and inverse Compton cooling off the CMB
\citep{1986ApJ...301..522I}.  Metal enrichment and its effect
on cooling rates is ignored.  We identify dark matter haloes in the
output snapshots using the friends-of-friends algorithm.  In order to
calculate power spectra, we project the relevant particles onto a grid
to create a density field, using, in the case of gas particles, the
cloud-in-cell (CIC) scheme that accounts for the SPH kernel.

If a source population at redshift $z$ is assumed to have a line
emission comoving volume emissivity
$\epsilon(\nu_{\textup{obs}}(1+z))$, then the specific intensity of
the observed emission can be determined by solving the cosmological
radiative transfer equation. The angle-averaged solution at $z = 0$
can be written as
\begin{equation}
  I(\nu_\mathrm{obs}, z=0)=\frac{1}{4\pi} \int_{0}^{\infty}dz'\frac{dl}{dz'}
  \frac{\epsilon(\nu_\mathrm{obs}(1+z'))}{(1+z')^3},
\end{equation}
where $dl/dz = c/((1+z)H(z))$ denotes the proper line element, and we
have assumed that there is negligible absorption by the intervening
intergalactic medium.  Assuming an absence of contamination from other
redshifts, we can model the frequency dependence by a
$\delta$-function and write
\begin{equation}
  I(\nu_\mathrm{obs},
  z=0)=\frac{c}{4\pi}\frac{1}{H(z)}\frac{1}{\nu_\mathrm{em}(z)}
  \epsilon(\nu_\mathrm{obs}(1+z)),
  \label{eqn:i}
\end{equation}
where the $\nu_\mathrm{em}=\nu_{\textup{obs}}(1+z)$ is the rest-frame
emission frequency.

The volume-averaged emissivity $\epsilon$ is related to the line
luminosity $L$ of individual haloes by
\begin{equation}
  \epsilon(z)=\int_{M_{\textup{min}}}^{\infty} dM \frac{dn}{dM} L_i(M,
  z),
\end{equation}
where $dn/dM$ is the halo mass function.  $M_\mathrm{min}$ is the
minimum mass of haloes that can form stars and produce line emission.
At $z=7$, the minimum halo mass in our simulation is $2.3\times
10^{8}$~$h^{-1}$M$_\odot$, which is close to the atomic hydrogen
cooling limit.  The maximum halo mass at this redshift is $3.1\times
10^{12}$~$h^{-1}$M$_\odot$.  In order to model the emissivity
$\epsilon(z)$, we now need to model the halo luminosities $L(M, z)$.

\subsection{Star formation rate}

Linking the halo luminosities $L(M, z)$ to the star formation rate
(SFR) can be done either using observational data or theoretical
models of the emission processes of the different lines. The mechanism
of line emission is complex; it depends on, e.g., the morphology and
structure of galaxies, their metallicity, radiation field and
density. Line emission can be excited by starlight, dissipation of
mechanical energy by turbulence and shocks, or by the active galactic
nuclei.  In the reionization epoch, CMB heating and attenuation can
also be important \citep{2017arXiv171100798L}. Several empirical
models have been proposed for the emission of different lines, e.g.,
CO \citep{2009ApJ...698.1467O, 2011ApJ...728L..46G}, \lya
\citep{2013ApJ...763..132S, 2014ApJ...786..111P,2017ApJ...846...21F},
C~\textsc{ii} \citep{2012ApJ...745...49G, 2016ApJ...833..153S}, but
all of them rely on sets of poorly known parameters that characterize
the galaxies and their interstellar medium in the reionization era.
For the C~\textsc{ii} line, while individual galaxies have been
detected at $z>6$ \citep[e.g.,][]{2016MNRAS.462L...6K,
  2016ApJ...829L..11P, 2017ApJ...836L...2B, 2017A&A...605A..42C,
  2017ApJ...842L..15S}, a complete understanding of the line
excitation is still lacking.

\begin{figure}
  \begin{center}
    \includegraphics[width=\columnwidth]{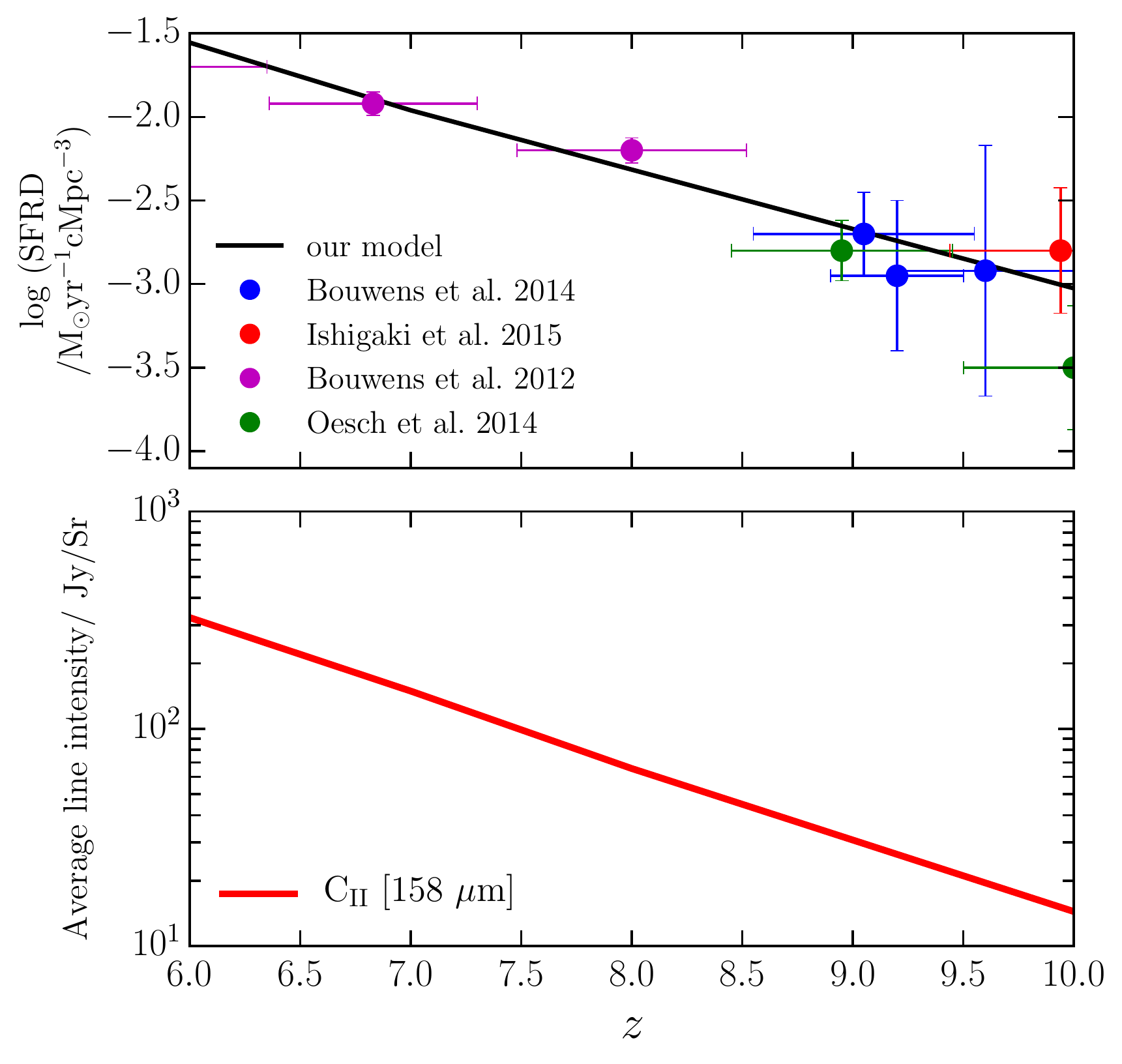}
  \end{center}
  \caption{Top panel shows the star formation rate density evolution
    in our model (black curve) in comparison with various
    extinction-corrected observational measurements (coloured
    symbols).  Bottom panel shows the resultant evolution of the
    average intensity of \c2 line emission.}
  \label{fig:SFR_plot}
\end{figure}

Considering the large amount of uncertainties in the detailed
modelling, we will continue our study using empirical relations from
the literature that relate the halo luminosity to its star formation
rate (SFR) as a power law, 
\begin{equation}
  L_i\propto \textup{SFR}^\gamma,
\end{equation}
where the exponent $\gamma$ encodes possible nonlinearities due to
processes such as collisional excitation \citep{2017arXiv171100798L}.
We assume that the SFR of a halo of mass $M$ is proportional to the
halo mass
\begin{equation}
  \mathrm{SFR}=f_*(z)M_\mathrm{halo},
\end{equation}
where we obtain the redshift-dependent proportionality factor by
assuming a linear evolution of the $\log(\mathrm{SFRD})$ with redshift
and calibrating $f_*(z)$ so that the resultant SFR density in the
simulation box is consistent with observed
data \citep{2015ApJ...808..104O}.  Figure~\ref{fig:SFR_plot} shows the
SFR density in our model in comparison with extinction-corrected
observational measurements.

\subsection{CII line emission}

Once we have the star formation rate model, we assign \c2 line
luminosities $L_\mathrm{[CII]}$ to each halo in our simulation box, by
using the predicted L$_\mathrm{[CII]}$--SFR relation from the model
presented by \cite{2017arXiv171100798L},
\begin{multline}
\label{eq:CII-SFR}
\log\left(\frac{\mathrm{L_{\mathrm{[CII]}}}}{L_{\odot}}\right) = (1.4
- 0.07z)\times \log\left(\frac{\mathrm{SFR}}{\mathrm{M}_{\odot}
  \mathrm{yr}^{-1}}\right)\\ + 7.1- 0.07z \,.
\end{multline}
In this paper, the semi-analytical model (SAM) of galaxy formation
\textsc{g.a.s.} described in \cite{2015A&A...575A..32C,
  2016A&A...589A.109C} was used, after further modifications assuming
an inertial turbulent cascade in the gas that generates a delay
between the accretion of the gas and the star formation (Cousin and
Guillard, submitted).  It is assumed that the \c2 emission in high-$z$
galaxies arises predominantly from photo-dominated regions (PDR). For
each galaxy in the SAM, an equivalent PDR characterised by three
parameters (the mean hydrogen density, gas metallicity, and
interstellar radiation field) is defined. The \c2 line emission is
then computed using the \textsc{cloudy} photoionisation code
\citep{C17}. This model allows computation of the \c2 luminosity for a
large number of galaxies (e.g., 28,000 at $z=5$).  It takes into
account the effects of CMB heating and attenuation that are important
at such high redshifts.  The model is able to reproduce the
$\mathrm{L_{\mathrm{[CII]}}}$--SFR relation observed for 50
star-forming galaxies at $z\ge 4$. We used here the mean relation
given in Equation~(\ref{eq:CII-SFR}) although it is found that the
$\mathrm{L_{\mathrm{[CII]}}}$--SFR relation is very dispersed (0.51 to
0.62 dex from $z=7.6$ to $z=4$). The large dispersion is due to the
combined effect of different interstellar radiation fields,
metallicities, and gas contents in the simulated high-redshift
galaxies.

\begin{figure*}
  \begin{center}
    \includegraphics[width=0.8\textwidth]{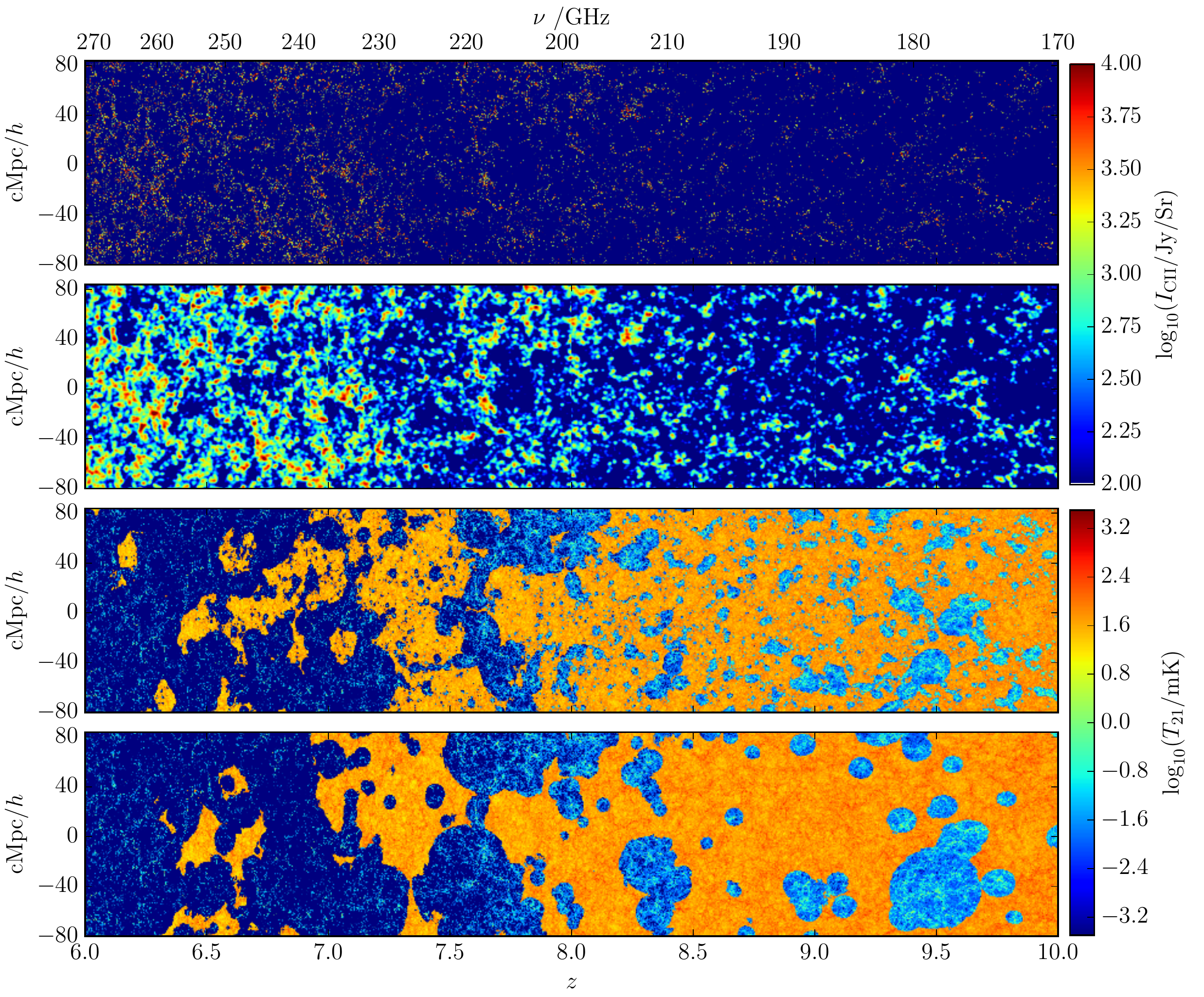} 
  \end{center}
  \caption{Light cones of \c2 and 21~cm brightness from redshift $z=6$
    to $10$.  The top panel shows individual sources corresponding to
    individual halos with mass $M_\mathrm{min}=10^8 M_\odot$; the
    bottom panel shows the \c2 intensity map with a resolution of
    $0.4^\prime$ ($\sim 1$~cMpc at $z=6$).  The middle panel shows the
    21~cm light cone in the reionization model with low-mass sources,
    while the bottom panel shows the 21~cm light cone in the
    reionization model with high-mass sources.  The anti-correlation
    between the \c2 and 21~cm maps is visually apparent in the
    low-mass reionization model.}
    \label{fig:lightcone}
\end{figure*}

In order to calculate the three-dimensional distribution of the
specific \c2 line intensity, we created coeval emission maps by
assigning to each halo in the simulation volume a line luminosity
$L_\mathrm{[CII]}(M)$ modelled as above.  Using the information about
their spatial positions, we then sum the volume emissivities in each
cell of a uniform $512^3$ grid to obtain three-dimensional emission
maps representing comoving regions of space of volume
$(160~\mathrm{cMpc}/h)^3$.  Using Equation~(\ref{eqn:i}), the observed
specific intensity corresponding to the cell is then given by
\begin{equation}
  I_\mathrm{cell} = \frac{c}{4\pi} \frac{1}{\nu_\mathrm{[CII]}H(z)}
  \frac{L_\mathrm{[CII],cell}}{V_{\textup{cell}}},
  \label{eqn:icell}
\end{equation}
where $L_\mathrm{[CII],cell}$ is the luminosity of the cell, given by
the sum of the luminosities of any haloes located in the cell.
Figure~\ref{fig:SFR_plot} shows the evolution of the average line
intensity.  Figure~\ref{fig:lightcone} shows a light cone of the \c2\
specific intensity created by interpolating between simulation
snapshots spaced at 40~Myr intervals between $z=6$ and $10$.  The
simulation corresponds to a total survey area of about $1.5\times 1.5$
deg$^2$, with each cell occupying an area of $0.2'\times 0.2'$.  At
redshift $z=7$, a comoving distance of $160~\mathrm{cMpc}/h$ along the
observation axis corresponds to about $\Delta z=0.5$.

\begin{figure*}
  \begin{center}
    \includegraphics[width=0.85\textwidth]{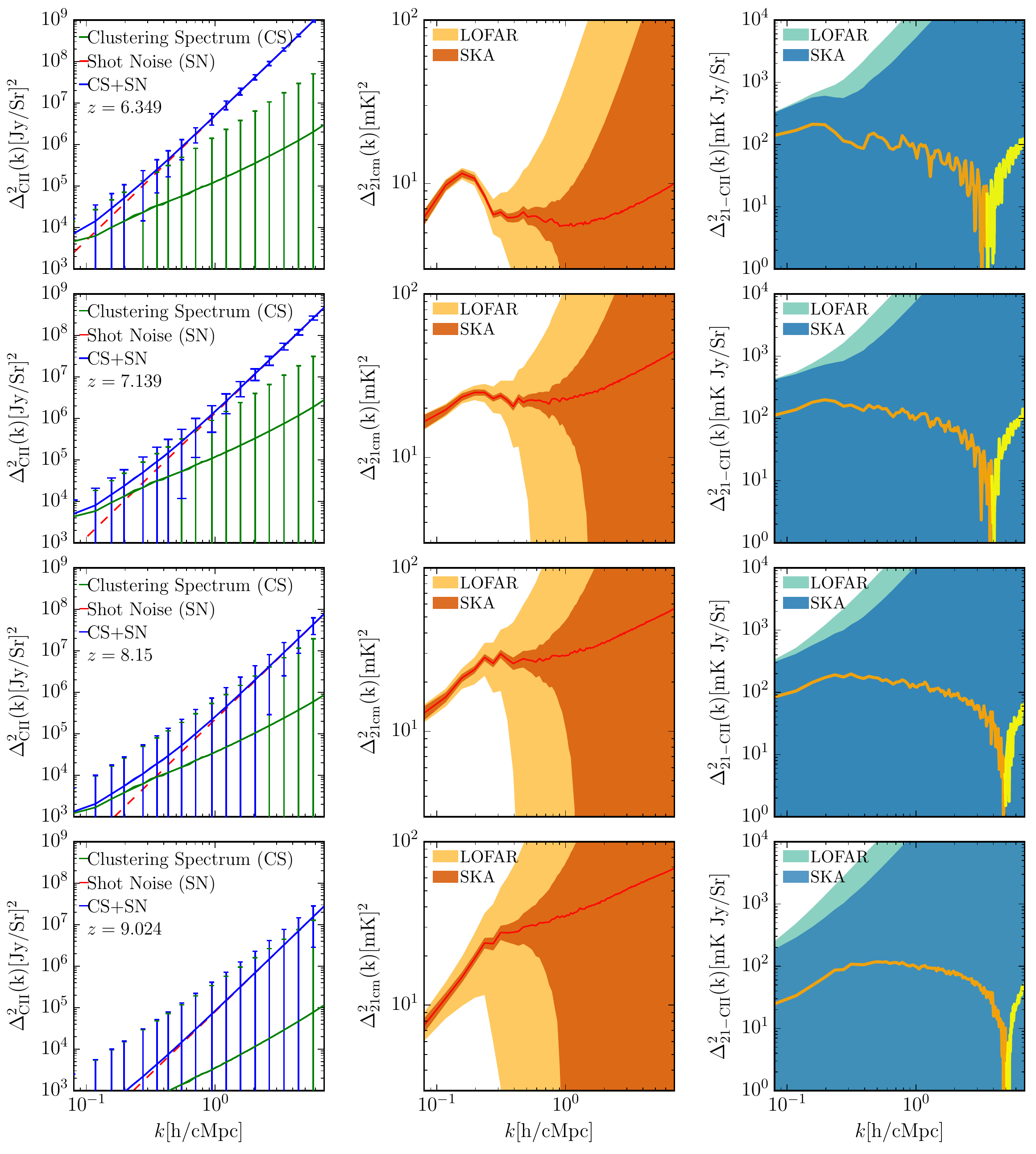}  
  \end{center}

  \caption{The \c2 power spectrum, 21~cm power spectrum, and \c2-21cm
    cross power spectrum at redshifts $z=6$--$9$ for the
    reionization model with low-mass sources.  Red dashed and green solid
    curves in the left column show the shot noise and clustering
    contribution to the power spectrum, respectively.  The blue curves
    in this column shows the total power spectrum.  Error bars on
    the \c2 power spectra show the 1$\sigma$ sensitivities for
    CONCERTO for $\Delta z=0.5$ at $z = 7$, relative to the total
    power spectrum in blue and relative to the clustering power
    spectrum in green.  Two sets of shaded regions show errors
    corresponding to LOFAR and SKA1-LOW.  On the cross power spectra
    on the right panels, orange (yellow) lines are for negative
    (positive) cross-correlation coefficients.}

\label{fig:ps_gal}
\end{figure*}

\begin{figure*}
  \begin{center}
    \includegraphics[width=0.85\textwidth]{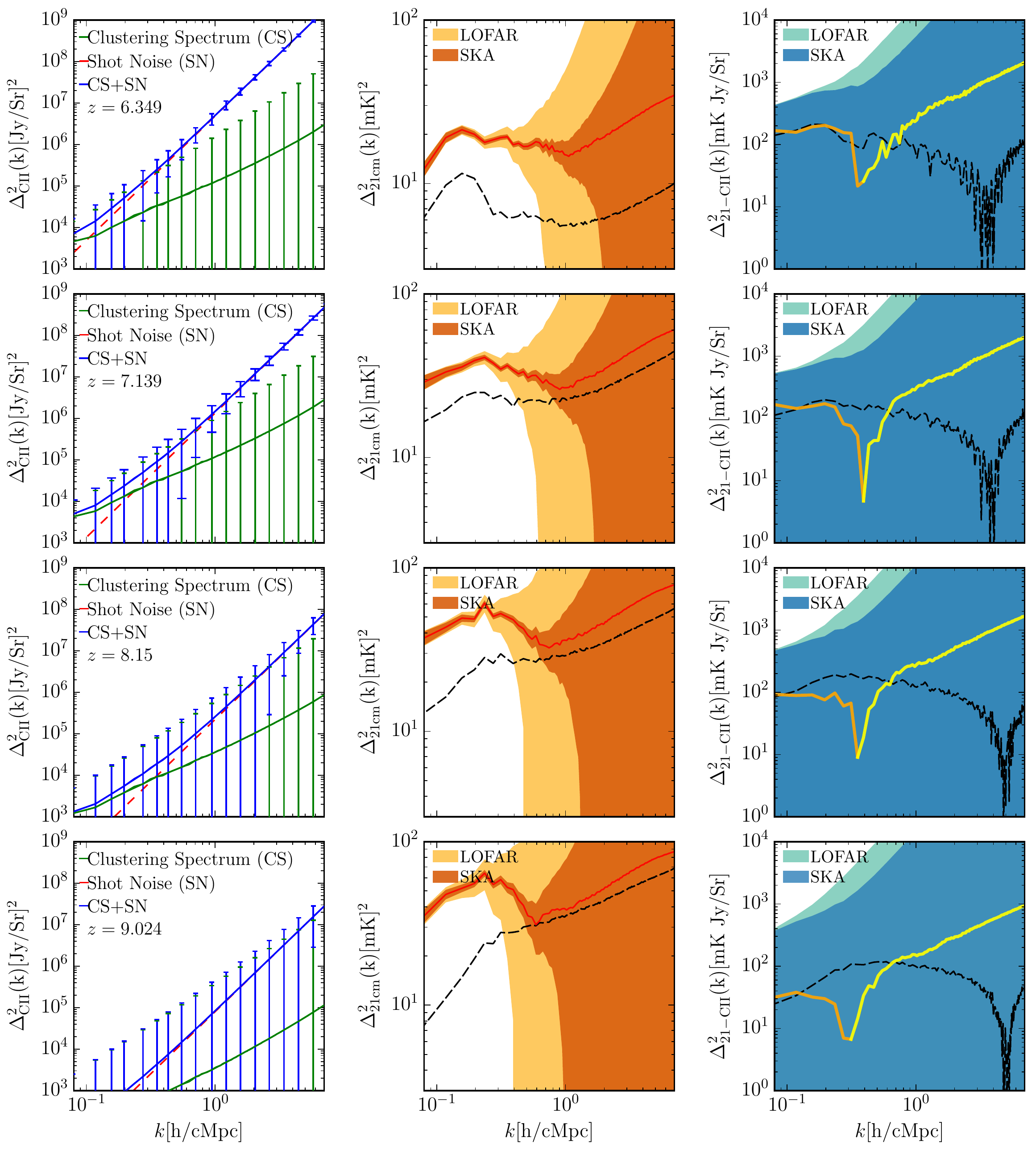}  
  \end{center}
  \caption{As Figure~\ref{fig:ps_gal} but for the 
    reionization model with high-mass sources.  The \c2 power spectra are identical to those
    in Figure~\ref{fig:ps_gal}.  The 21~cm power spectra and the
    \c2-21cm cross power spectra from Figure~\ref{fig:ps_gal} are
    shown in dashed black for comparison.}
  \label{fig:ps_agn}
\end{figure*}

\begin{figure}
  \includegraphics[width=0.5\textwidth]{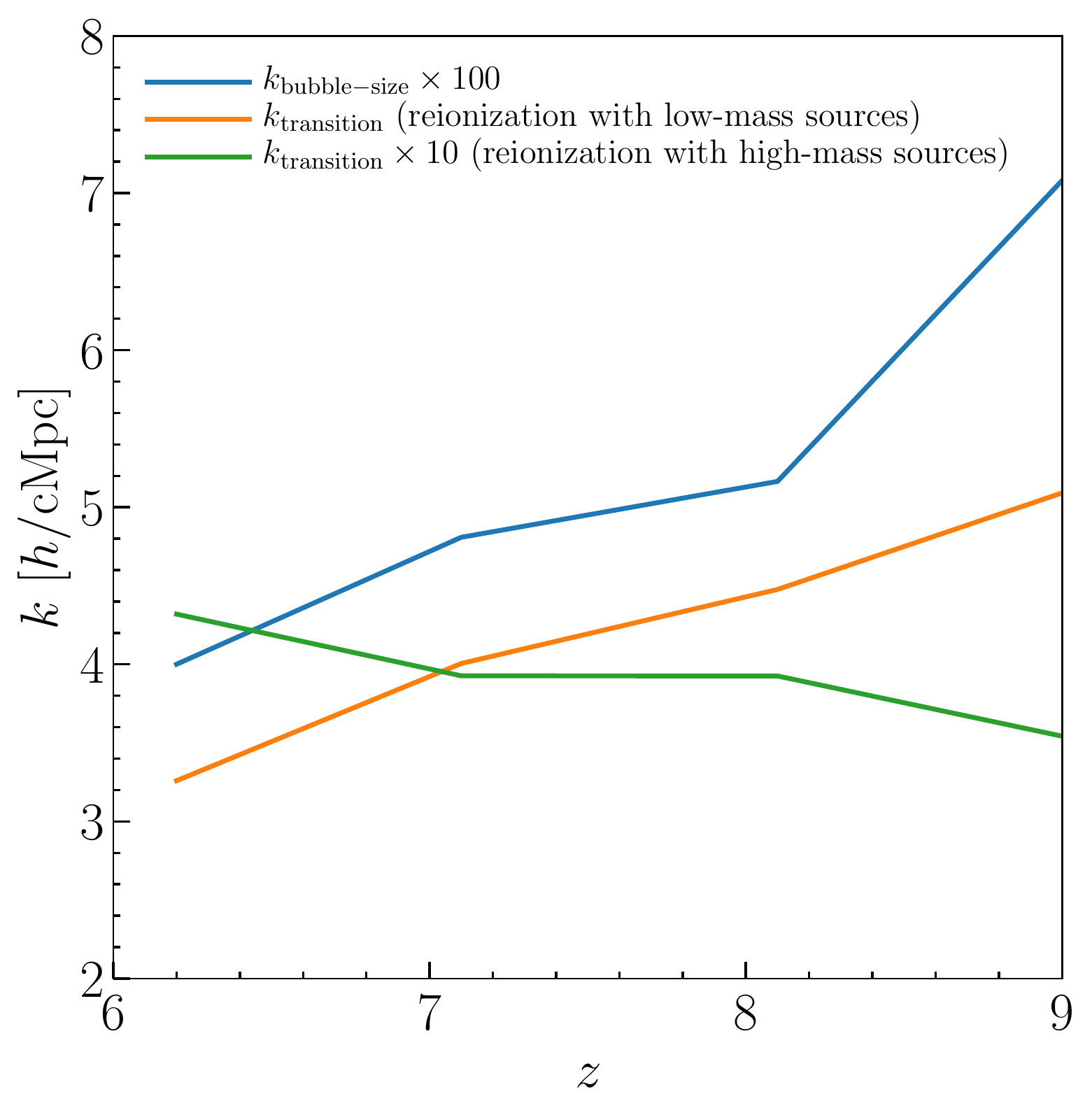} \caption{Evolution
  of the scale at which the \c2-21cm cross power spectrum transitions
  from negative to positive values in the reionization model with
  low-mass sources (orange curve) and the reionization model with
  high-mass sources (green curve).  The blue curve shows the evolution
  of the average ionized bubble size, defined simply at the the cube
  root of the ionized volume in the simulation box.  The transition
  scale tracks the bubble size evolution in the model with low-mass
  sources, but not in the model with high-mass sources.  The
  quantities $k_\mathrm{bubble-size}$ and $k_\mathrm{transition}$ for
  the high-mass reionization model have been multiplied by factors of
  100 and 10, respectively, for easier
  comparison.}  \label{fig:bubblesize}
\end{figure}

\subsection {Power spectra}
\label{sec:clustering}

We derive three-dimensional spherically-averaged power spectra of the
\c2\ line emission in our model as
\begin{equation}
  \Delta^2(k) = \frac{k^3}{2\pi^2}\cdot
  \frac{\langle\widetilde{I}^2(k)\rangle}{V_\mathrm{box}},
  \label{eqn:ps}
\end{equation}
where $\widetilde{I}$ is the Fourier transform of the specific
intensity defined in Equation~(\ref{eqn:i}), and $V_\mathrm{box}$ is
the box volume, $(160\ $cMpc$/h)^3$.  We ignore the anisotropies
arising from redshift-space distortions and the redshift evolution
across the box.  Left column of Figure~\ref{fig:ps_gal} shows the
resultant power spectra for the \c2\ emission for redshifts from $z=6$
to $9$.  The shot noise contribution to the \c2\ power spectrum is
included.  The shot noise is given by
\begin{equation}
  \Delta^2_\mathrm{shot}(k, z) =
  \frac{k^3}{2\pi^2}\left[\frac{c}{4\pi\nu_\mathrm{[CII]}
      H(z)}\right]^2\sum_i \frac{\left[L_\mathrm{CII}(M_i,
      z)\right]^2}{V_\mathrm{box}},
\end{equation}
where $L_\mathrm{CII}(M_i, z)$ is the C~\textsc{ii} luminosity (in
erg\ s$^{-1}$) of halo $i$ with mass $M_i$, and the summation is over
all haloes.  The frequency $\nu_\mathrm{CII}$ is the rest-frame
frequency of the C~\textsc{ii} line.  Shot noise dominates the power
spectrum at $k\gtrsim 0.5$~\kunit\ \citep{2016ApJ...833..153S}, but is
irrelevant in the \c2-21cm cross power spectrum discussed in this
paper, as the 21~cm emission comes from the extended IGM.  Also shown
in Figure~\ref{fig:ps_gal} are the sensitivities corresponding to
experimental configurations, which we discuss below.

The line emission power spectra trace the halo power spectrum, with a
constant bias factor as the emission amplitude is simply proportional
to the halo mass.  The amplitude of the \c2 power spectrum decreases
from redshift $z=6$ to 9 by a factor of 100.  Our values are
consistent with those from other models in the literature
\citep{2016ApJ...833..153S, 2015ApJ...806..209S, 2012ApJ...745...49G}.

\section{21\,cm line maps}
\label{sec:HI_model} 

The redshifted 21~cm signal originates in the neutral intergalactic
regions.  We model the brightness temperature at location $\mathbf{x}$
in our simulations in a similar manner as \citet{2017MNRAS.469.4283K}
\begin{equation}
  T_b(\mathbf{x})=\overline T_b
  x_\mathrm{HI}(\mathbf{x})\Delta(\mathbf{x}),
  \label{eqn:tb}
\end{equation}
where the mean temperature $\overline T_b\approx 22 \mathrm{mK}
[(1+z)/7]^{1/2}$ \citep{2009MNRAS.394..960C}, $x_\mathrm{HI}$ is the
neutral hydrogen fraction in a cell, and $\Delta$ is the gas density
in units of the average density in the simulation.  We neglect the
impact of redshift space distortions due to peculiar velocities.  We
also assume that the spin temperature is much greater than the CMB
temperature and that the Ly$\alpha$ coupling is sufficiently complete
throughout the IGM.

We derive the ionization field by placing sources of Lyman-continuum
radiation in dark matter haloes and using the well-known excursion set
method \citep{2004ApJ...613...16F, 2009MNRAS.394..960C,
2011MNRAS.411..955M}.  The total number of ionizing photons
$N_{\gamma}$ produced by a halo is assumed proportional to the halo
mass \citep{2016MNRAS.463.2583K}
\begin{equation}
  N_\gamma(M) = N_\gamma^\mathrm{LyC}M,
  \label{eqn:ngammam}
\end{equation}
where the proportionality factor $N_\gamma^\mathrm{LyC}$ includes the
Lyman-continuum escape fraction.  A grid cell at position $\mathbf{x}$
is ionized if the condition
\begin{equation}
  \langle n_\gamma(\mathbf{x})\rangle_R > \langle n_\mathrm{H}(\mathbf{x})\rangle_R(1+\bar N_\mathrm{rec}),
  \label{eqn:exset1}
\end{equation}
is satisfied in a spherical region centred on the cell for some radius
$R$ \citep{2004ApJ...613....1F, 2009MNRAS.394..960C,
  2011MNRAS.411..955M}.  Here, the averages are over the spherical
region, $n_\mathrm{H}$ is the hydrogen number density, 
\begin{equation}
  n_\gamma = \int_{M_\mathrm{min}}^\infty dM\left.\frac{dN}{dM}\right\vert_{R}N_\gamma(M),
  \label{eqn:ngamma}
\end{equation}
where $dN/dM|_R$ is the halo mass function within the spherical
region, $M_\mathrm{min}$ is the minimum halo mass that contributes
ionizing photons, and $\bar N_\mathrm{rec}$ is the average number of
recombinations per hydrogen atom in the IGM.  The condition in
Equation~(\ref{eqn:exset1}) can be recast as
\begin{equation}
  \zeta_\mathrm{eff}f(\mathbf{x},R)\geq 1,
  \label{eqn:exset}
\end{equation}
where the quantity
\begin{equation}
  f=\rho_m(R)^{-1}\int_{M_\mathrm{min}}^\infty
  dM\left.  \frac{dN}{dM}\right\vert_{R}M, \label{eqn:fcoll}
\end{equation}
is the collapsed fraction into haloes of mass $M>M_\mathrm{min}$,
$M_\mathrm{min}$ is the minimum mass of halos that emit Lyman
continuum photons, $\rho_m(R)$ is the average matter density, and
$dN/dM|_R$ is the halo mass function in the sphere of radius $R$.  The
parameter $\zeta_\mathrm{eff}$ quantifies the number of photons in the
IGM per hydrogen atom in stars, accounting for hydrogen recombinations
in the IGM.  We can write $\zeta_\mathrm{eff}$ in terms of the
parameters of Equations~(\ref{eqn:ngammam}) and (\ref{eqn:exset1}) as
\begin{equation}
  \zeta_\mathrm{eff} = \frac{N^\mathrm{LyC}_\gamma}{1-Y_\mathrm{He}}(1+\bar N_\mathrm{rec})^{-1},
\end{equation}
where $Y_\mathrm{He}$ is the helium mass fraction.  This is the only
parameter that determines the ionization field in this approach.  The
volume-weighted ionized fraction in the simulation box is
$Q_V\equiv \sum_iQ_i/n_\mathrm{cell}$, where the ionized volume
fraction in a cell $i$ is $Q_i$ and $n_\mathrm{cell}$ is the total
number of grid cells.

We consider two reionization models in this paper.  The evolution of
the ionized fraction $Q_V$ is identical in both models, and follows
the evolution in the Late/Default model of \cite{2016MNRAS.463.2583K}.
This is achieved by solving for $\zeta_\mathrm{eff}$ for the assumed
$Q_V$.  The simulation box is completely ionized, i.e., $Q_V=1$, at
$z=6$.  This evolution of the ionized fraction is consistent with the
constraint from the CMB measurement of the electron scattering optical
depth.  The two models differ however in the range of halo masses that
contribute to reionizing photons.  In one of the models, we set the
value of the minimum halo mass in Equation~(\ref{eqn:fcoll}) to be
$M_\mathrm{min}=2.3\times 10^{8}$~M$_\odot$, which is approximately
the mass of the smallest halo resolved in our simulation at $z=7$.
This model should represent reionization dominated by star-forming
galaxies reasonably well.  In our second reionization model, we assume
$M_\mathrm{min}=10^{11}$~M$_\odot$.  Only high-mass haloes contribute
to reionization in this model.
These reionization models with low-mass and high-mass sources 
present two plausible but distinct cases of source clustering, which
is the quantity of interest that we want to explore later in this
paper by studying its effect on the 21~cm power spectrum and
the \c2-21cm cross power spectrum.

The bottom two panels of Figure~\ref{fig:lightcone} show the evolution
of 21\,cm brightness in our two reionization models.  These light
cones are analogous to those obtained for the \c2 emission, shown in
the top two panels of this figure.  Although the average ionized
hydrogen fraction is the same in the two reionization models, the
distribution of the 21~cm signal is quite different.  The reionization
model with high-mass sources has large and more clustered ionized
regions with low 21\,cm brightness.  More importantly, in the
reionization model with low-mass sources, every source of \c2 emission is
also a source of hydrogen-ionizing photons.  As a result, the
distribution of the 21\,cm signal is anti-correlated with that of the
\c2 signal: every \c2 source is located in regions with low 21\,cm
brightness.  In the reionization model with high-mass sources, on the
other hand, \c2 emitters in haloes with masses less than
$M_\mathrm{min}=10^{11}$~M$_\odot$ do not contribute any
hydrogen-ionizing photons.  As a result, these low-mass \c2-emitters
are located in neutral regions, which are bright in 21\,cm. This has
an important effect on the \c2-21cm correlation.

The middle columns of Figures~\ref{fig:ps_gal} and \ref{fig:ps_agn}
show the predicted 21\,cm power spectra in our simulation in the
reionization models with low-mass and high-mass sources, respectively.
The power spectrum has a familiar shape: at small scales it is
dominated by the matter power spectrum, and at large scales by a
prominent ``bump'' due to ionized bubbles.  At $k=0.1~h/$cMpc the
amplitude of the 21\,cm power spectrum evolves from $\Delta^2(k)\sim
2$~mK$^2$ at $z=9$ to 10~mK$^2$ at $z=7.1$ in the reionization model
with low-mass sources.  In the high-mass case, the large-scale
amplitude of the 21~cm power spectrum is higher, with $\Delta^2(k)\sim
35$~mK$^2$ at $z=9$ to 30~mK$^2$ at $z=7.1$, due to the higher
clustering of ionized regions \citep{2017MNRAS.469.4283K}.
Figures~\ref{fig:ps_gal} and \ref{fig:ps_agn} also show the
sensitivity of experiments aiming to detect the 21\,cm signal.  We
discuss this in Section~\ref{sec:im} below.

\section{The [C~II]-21cm cross power spectrum}
\label{sec:cross-corr}

An exciting prospect for high-redshift \c2 intensity mapping is to
combine it with observations of the coeval redshifted 21\,cm line
signal from the epoch of reionization.  A detection of the \c2-21cm
cross power spectrum will assist in foreground decontamination and
complement the \c2 and 21~cm power spectra as a probe of the epoch of
reionization \citep{2010JCAP...11..016V, 2011ApJ...741...70L}.
Furthermore, the \c2-21cm cross power spectrum may
act as a direct tracer of the growth of ionizing bubbles during
reionization \citep{2012ApJ...745...49G}.

As discussed above, Figure~\ref{fig:lightcone} shows light cones of
the \c2 and 21\,cm intensity.  Typically, on large scales, we expect
the \c2 emission from halos and the 21\,cm signal from the IGM to be
anti-correlated, because fully neutral regions do not contain emitting
galaxies, while the halo-rich regions are depleted of neutral
hydrogen.  On scales smaller than the ionized bubbles, however, there
is positive correlation between the two fields.  This behaviour is
visually apparent in Figure~\ref{fig:lightcone}, particularly at
redshift $z\sim 7$, where the ionized regions are sufficiently large.

In order to study this cross-correlation quantitatively, we define the
cross power spectrum of the \c2 and 21\,cm intensity maps as
\begin{equation}
  \Delta^2(k) = \frac{k^3}{2\pi^2} \cdot \frac{1}{V_\mathrm{box}}
  \cdot \frac{\langle\widetilde{I_1}^*(k)\widetilde{I_2}(k) +
    \widetilde{I_1}(k)\widetilde{I_2}^*(k)\rangle}{2},
    \label{eqn:crossps}
\end{equation}
where $I_1$ and $I_2$ denote the intensities of \c2 and 21~cm,
respectively.  The quantity $\tilde I$ is the Fourier transform of
$I$, and $\tilde I^*$ is the complex conjugate of $\tilde I$. The
result is shown in the right column of Figures~\ref{fig:ps_gal}
and \ref{fig:ps_agn} for our reionization models with low-mass and
high-mass sources, respectively.  On large scales the
cross-correlation is negative, as expected.  In both models, at
$k=0.1$~\kunit, the value of the cross power spectrum is $\sim
10^2$~mK~Jy$/$Sr at redshift $z\sim 9$.  This increases to close to
$5\times 10^2$~mK~Jy$/$Sr at $z\sim 6$.  (Figures~\ref{fig:ps_gal}
and \ref{fig:ps_agn} also show the experimental sensitivities for
measuring the cross power spectra; we discuss this in the next
section.)

The scale at which the cross power spectrum transitions from positive
to negative values is quite different in the two reionization models.
In the low-mass model, this scale is at
$k_\mathrm{transition}=3$--$5$~\kunit, while it is close to
$k_\mathrm{transition}=0.3$~\kunit\ in the high-mass model.
This is consistent with the picture that the transition scale measures
the average size of ionized regions.  As seen in
Figure~\ref{fig:lightcone}, the ionized regions are larger in the
high-mass model, which is reflected in the value of the
transition scale of the cross power spectrum.
Figure~\ref{fig:bubblesize} shows the evolution of the transition
scale in the two models.  The blue curve in this figure shows the
evolution of the average bubble size $k_\mathrm{bubble-size}$ in the
simulation, as measured by the cube root of the ionized volume.  The
evolution of the cross power spectrum transition scale in the galaxy
dominated model follows that of $k_\mathrm{bubble-size}$, whereas the
evolution in the transition scale for the high-mass model has a
qualitatively different trend.  This is because in the
reionization model with low-mass sources, each \c2\ source is also a
source of hydrogen ionizing photons.  Therefore, every \c2\ source is
in an ionized region, and there is perfect anti-correlation between
the \c2\ and 21\,cm fields at scales larger than the bubble size.
This is not the case in the high-mass reionization model, where
most \c2\ sources lie in neutral regions.

\begin{figure*}
  \begin{center} \includegraphics[width=0.85\textwidth]{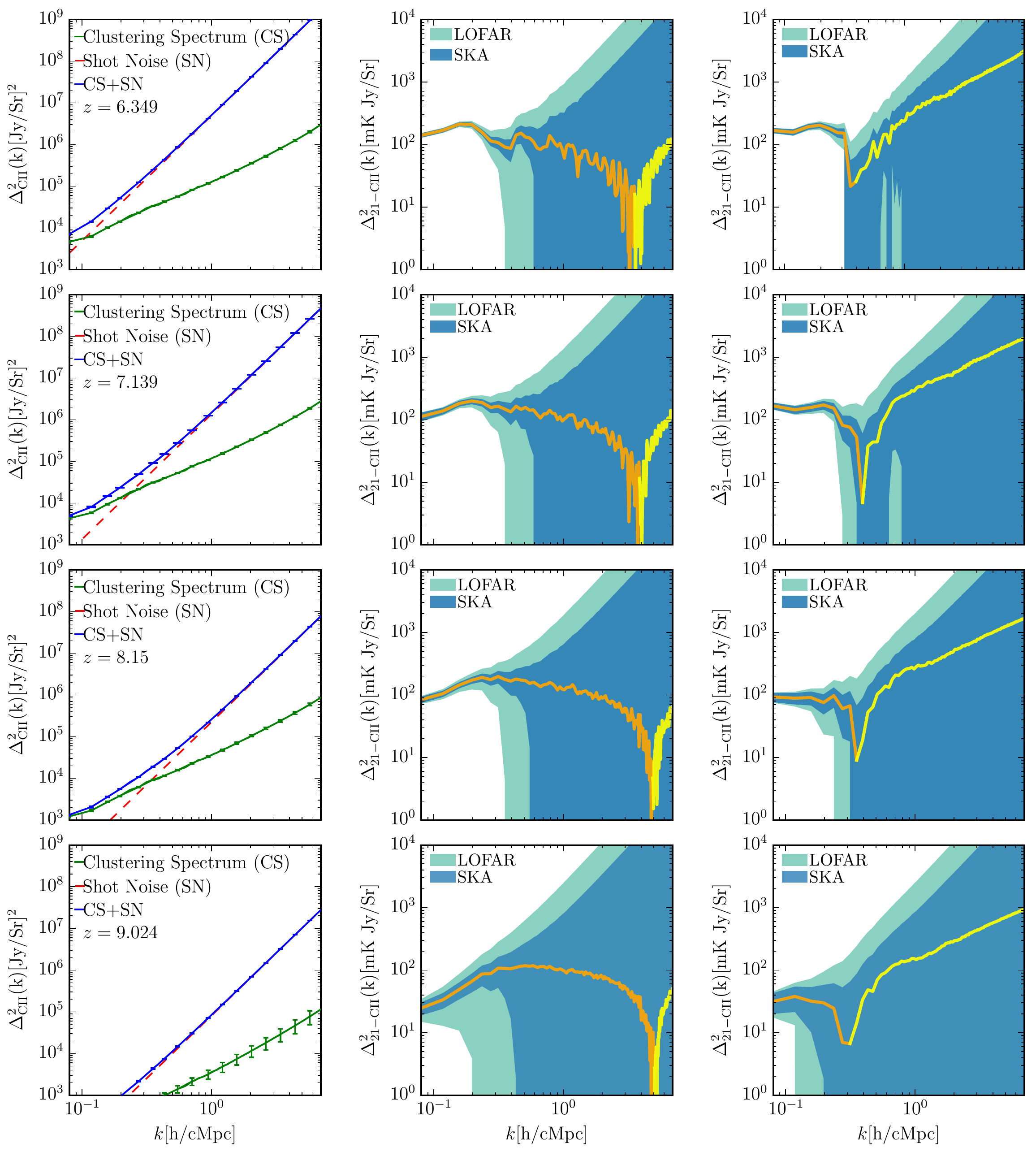} \end{center} \caption{The \c2
    power spectrum, and the \c2-21\,cm cross power spectrum for
    reionisation models with low-mass and high-mass sources at
    redshifts $z=6$--$9$.  Errorbars on the \c2 power spectra show
    sensivities for the Stage~II experiment assuming a survey area of
    $10~\mathrm{deg}^2$ and integration time of 1000~hr.  Two sets of
    shaded regions on the cross power spectra show errors
    corresponding to LOFAR and SKA surveys.}  \label{fig:ps_gal_ccat}
\end{figure*}

\section{Intensity mapping experiments}
\label{sec:im}

To estimate the feasibility of \c2 intensity mapping, we consider the
the CONCERTO experiment \citep{2018arXiv180108054L,
  2016ApJ...833..153S}.  We also consider a successor Stage~II
experiment beyond CONCERTO.  The specifications for these two
experiments are summarised in Table~\ref{tab:params}.  Our choice of
the Stage~II experiment parameters is inspired by the CCAT-p
telescope\footnote{www.ccatobservatory.org/docs/pdfs/Draft\_CCAT-p.prospectus.170809.pdf}.
For the 21~cm signal from the same redshifts, we consider measurements
using LOFAR and SKA.

\subsection{[C~II] experimental sensitivities}
\label{sec:c2sens}

We estimate the sensitivity of experiments to measure the \c2 power
spectrum by computing the uncertainty on the power spectrum following
\cite{2011ApJ...741...70L, 2012ApJ...745...49G} and
\cite{2016ApJ...833..153S}:
\begin{equation}
  \mathrm{var}[P_\mathrm{CII}(k)]=\frac{[P_\mathrm{CII}(k)+P_\mathrm{CII}^\mathrm{N}(k)]^2}{N_\mathrm{m}(k,z)}, 
  \label{eqn:psens}
\end{equation}
where $P_\mathrm{CII}(k)$ is the model power spectrum, $N_\mathrm{m}$
is the number of modes in the survey volume with wavenumber $k$ at
redshift $z$, and $P_\mathrm{CII}^\mathrm{N}$ is the noise power
spectrum.  The noise power spectrum is given by
\begin{equation}
  P_\mathrm{CII}^\mathrm{N} =
  V_\mathrm{pix}\frac{\sigma^2_\mathrm{pix}}{t_\mathrm{pix}},
  \label{eqn:noiseps}
\end{equation}
where $V_\mathrm{pix}$ is the volume surveyed by a single pixel,
$t_\mathrm{pix}$ is the observing time per pixel, and
$\sigma^2_\mathrm{pix}$ is the noise variance per spectral element.
The observing time per pixel is given by
\begin{equation}
  t_\mathrm{pix}=t_\mathrm{survey}N_\mathrm{pix}\frac{\Omega_\mathrm{beam}}{A}. 
\end{equation}
Here, $t_\mathrm{survey}$ is the survey duration, which we take to be
1500~hr.  The beam area $\Omega_\mathrm{beam}$ is given by
$\Omega_\mathrm{beam}=2\pi\left(\theta_\mathrm{beam}/2.355\right)^2$,
where $\theta_\mathrm{beam}=1.22\lambda_\mathrm{obs}/D$ and $D=12$\,m
for CONCERTO.  We assume a survey area of $A=2$~deg$^2$.  The volume
surveyed by one pixel is given by \citep{2012ApJ...745...49G}
\begin{multline}
V_\mathrm{pixel}(z) = 1.1\times 10^3 (\mathrm{cMpc}/h)^3
\left(\frac{\lambda}{158~\mu\mathrm{m}}\right)\\
\times \left(\frac{1+z}{8}\right)^{1/2}
\left(\frac{\theta_\mathrm{beam}}{10~\mathrm{arcmin}}\right)^2
\left(\frac{\delta\nu}{400~\mathrm{MHz}}\right).
\end{multline}
The noise variance $\sigma^2_\mathrm{pix}$ in Equation~(\ref{eqn:noiseps}) is given by 
\begin{equation}
  \sigma^2_\mathrm{pix}=\frac{\mathrm{NEI}_\mathrm{diff}^2}{N_\mathrm{pix}},
\end{equation}
where the noise equivalent power input from diffuse emission, defined
as the power from diffuse emission absorbed that produces a
signal-to-noise ratio of unity at detector output, is (in
MJy~sr$^{-1}$~s$^{1/2}$)
\begin{equation}
  \mathrm{NEI}_\mathrm{diff}=\mathrm{NEI}\times\frac{10^{-9}}{\Omega_\mathrm{beam}}.
\end{equation}
For CONCERTO, $\mathrm{NEI}/\sqrt{N_{pix}}$=155 mJy~s$^{1/2}$ (see
Table\,3 of \citealt{2016ApJ...833..153S}), assuming an overall
transmission of the system $T=0.3$, a spectral resolution
$\delta\nu=1.5$\,GHz, a number of pixel (and thus of spectrometer)
$N_{pix}$=1500, a precipitable water vapor of 2\,mm, an elevation of
60 degrees, and assuming the sensitivity already achieved by the NIKA2
KIDS detectors on sky \citep{2018A&A...609A.115A}.

The number of Fourier modes $N_\mathrm{m}$ in Equation~(\ref{eqn:psens}) is given by
\begin{equation}
  N_\mathrm{m}(k,z)=2\pi k^2\Delta k\frac{V_\mathrm{survey}}{(2\pi)^3}.
  \label{eqn:nmodes}
\end{equation}
Here, $\Delta k$ is the bin size assumed in $k$-space, and the survey
volume is given by
\begin{multline}
  V_\mathrm{survey}(z) = 3.7\times 10^7 (\mathrm{cMpc}/h)^3
  \left(\frac{\lambda}{158~\mu\mathrm{m}}\right) \\
  \times\left(\frac{1+z}{8}\right)^{1/2}
  \left(\frac{A}{16~\mathrm{deg}^2}\right)
  \left(\frac{B_{\nu}}{20~\mathrm{GHz}}\right).
\end{multline}
This allows us to estimate $\mathrm{var}[P_\mathrm{CII}(k)]$ using
Equation~(\ref{eqn:psens}).  (Note that
Equation~(\ref{eqn:nmodes}) is approximate and may lead to an
overestimated signal-to-noise ratio.)  Table~\ref{tab:params}
summarises all the properties of the CONCERTO experiment.

The left columns in Figures~\ref{fig:ps_gal} and \ref{fig:ps_agn} show
the uncertainties in the \c2 power spectrum for the CONCERTO
experiment from $z\sim 6$ to $\sim 9$.  We find that the CONCERTO
should be able to measure the large scale power spectrum of
\c2\ emission to redshifts of up to $z=8$ (with a signal-to-noise
ratio of $\sim 1$ at $k<0.1$~\kunit\ with 1500~hr of integration).
Our predictions thus agree with the ``pessimistic'' case discussed
by \citet{2018arXiv180108054L}.

For the Stage~II experiment, we consider a noise equivalent flux
density (NEFD) that is five times better than CONCERTO.  We assume an
aperture size of $D=6$~m, and a spectral resolution of
$\delta\nu=400$~MHz.  The survey duration is assumed to be
$t_\mathrm{survey}=1000$~hr, while the survey area is set to
$A=10$~deg$^2$.  These parameters are also summarised in
Table~\ref{tab:params}.  The left columns in
Figure~\ref{fig:ps_gal_ccat} shows the uncertainties in the \c2 power
spectrum for the Stage~II experiment from $z\sim 6$--$9$.  The
signal-to-noise ratio is now enhanced by a factor of $\sim 40$
relative to CONCERTO at $k= 0.2$~\kunit at $z=6$.  With the Stage~II
experiment, the power spectrum is detectable even at $z\sim 9$ with a
signal-to-noise ratio of $\sim 50$ at $k= 0.2$~\kunit.

\subsection{21~cm experimental sensitivities}
\label{sec:21cmsens}

We study here the detectability of the 21~cm power spectrum for Low
Frequency Array (LOFAR; \citealt{2013A&A...556A...2V}, and the low
frequency instrument from Phase 1 of the Square Kilometre Array
(SKA1-LOW; astronomers.skatelescope.org).  These are listed in
Table~\ref{tab:experiments}.  Similar to Equation~(\ref{eqn:psens}),
the variance of the power spectrum at mode $k$ and redshift $z$ is
given by
\begin{equation}
  \mathrm{var}[P_\mathrm{21}(k)]=\frac{[P_\mathrm{21}(k)+P_\mathrm{21}^\mathrm{N}(k)]^2}{N_\mathrm{m}(k,z)}.
  \label{eqn:p21sens}
\end{equation}
The noise power spectrum $P_\mathrm{21}^\mathrm{N}(k)$ is 
estimated similar to \citet{2012ApJ...753...81P}, and is given by 
\begin{multline}
  P_\mathrm{21}^\mathrm{N}(k)\approx
  X^2Y\frac{k^{-1/2}}{2\pi^2}\left(\frac{1}{B}\right)^{1/2}\left(\frac{1}{\Delta\ln
    k}\right)^{1/2}\\\times\frac{\Omega}{2t}T^2_\mathrm{sys}\frac{u_\mathrm{max}^{1/2}}{N}\frac{1}{\Omega^{1/4}}\frac{1}{t_\textrm{per-day}^{1/2}},  
  \label{eqn:thermalpower}
\end{multline}
where $u_\mathrm{max}$ is the maximum baseline $b_\mathrm{max}$ in
wavelength units, and $X$ and $Y$ are conversion factors from angles
and frequencies, respectively, to comoving distance
(See \citealt{2016MNRAS.463.2583K} for further details). We assume
$t_\textrm{per-day}=6$~hr for 120 days.  Also in
Equation~(\ref{eqn:thermalpower}), $N$ is the number of baselines and
$\Omega$ is the field of view of an element in the array.  The system
temperature is assumed to be \cite{2007isra.book.....T}
\begin{equation}
  T_\mathrm{sys}=60~\mathrm{K}\left(\frac{300~\mathrm{MHz}}{\nu_c}\right)^{2.25},
  \label{eqn:tsys}
\end{equation}
and calculate the thermal noise power for an integration over
120~days, assuming a bandwidth of 6~MHz, an observing time of 6~hr per
day for 120~days, and a mid-latitude location.

The middle columns of Figures~\ref{fig:ps_gal} and \ref{fig:ps_agn},
show the resultant uncertainties in the 21~cm power spectrum for LOFAR
(yellow) and SKA (brown).  These experiments are only sensitive to
large scales due to limited baselines.  Neither of the experiments are
sensitive to 21~cm power for $k\gtrsim 1$ cMpc$^{-1}h$.  SKA1-LOW has
much greater sensitivity than LOFAR primarily due to large number of
antenna elements.  The signal to noise ratio is about 100 for these
two experiments $k\sim 0.1$ cMpc$^{-1}h$.  LOFAR has sensitivity for
scales corresponding to $k\lesssim 0.2$ cMpc$^{-1}h$.  At $k\sim 0.1$
cMpc$^{-1}h$, the signal to noise ratio for LOFAR is $\sim 10$.

\begin{table*}
  \begin{center}
    \begin{tabular}{lll}
      \hline
      Parameter & CONCERTO & Stage~II \\
      \hline
      Aperture size ($D$) & 12~m & 6~m \\
      Transmission ($T$) & 0.3 & 0.3 \\
      Frequency window ($\Delta\nu$) & 80~GHz & 80~GHz \\
      Spectral resolution ($\delta\nu$) & 1.5~GHz & 0.4~GHz \\
      $\mathrm{NEI}/\sqrt{N_{pix}}$ & 155~mJy~s$^{1/2}$ & 31~mJy~s$^{1/2}$ \\
      Survey area ($A$) & 2~deg$^2$ & 10~deg$^2$ \\ 
      Survey duration ($t_\mathrm{survey}$) & 1500~hr & 1000~hr \\ 
      \hline
    \end{tabular}
    \caption{Specifications for \c2 experiments considered in this
      paper. \label{tab:params}}
  \end{center}
\end{table*}

\begin{table*}
  \begin{center}
    \begin{tabular}{lll}
      \hline
      Parameter & LOFAR & SKA1-LOW \\
      \hline
      Number of antennae ($N_\mathrm{ant}$) & 48 & 512 \\
      Effective collecting area ($A_\mathrm{eff}$) & 526.0~m$^2$ & 962.0~m$^2$ \\
      Maximum baseline ($b_\mathrm{max}$) & 3475.6~m & 40286.8~m \\
      Minimum baseline ($b_\mathrm{min}$) & 22.92~m & 16.8~m \\
      Survey duration per day ($t_\mathrm{per-day}$) & 6~hr & 6~hr \\
      Survey number of days & 120 & 120 \\
      System temperature ($T_\mathrm{sys}$) & Equation~(\ref{eqn:tsys}) & Equation~(\ref{eqn:tsys}) \\
      \hline
    \end{tabular}
  \end{center}
  \caption{Specifications for 21~cm experiments considered in this
    paper.  We use SKA parameters obtained by
    \citet{2016MNRAS.460..827G} which broadly agrees with the baseline
    distribution given in the latest SKA1-LOW configuration document
    (Document number SKA-SCI-LOW-001; date 2015-10-28;
    http://astronomers.skatelescope.org/documents/.)}
  \label{tab:experiments}
\end{table*}

\begin{figure*}
  \begin{tabular}{cc}
    \includegraphics[width=0.45\textwidth]{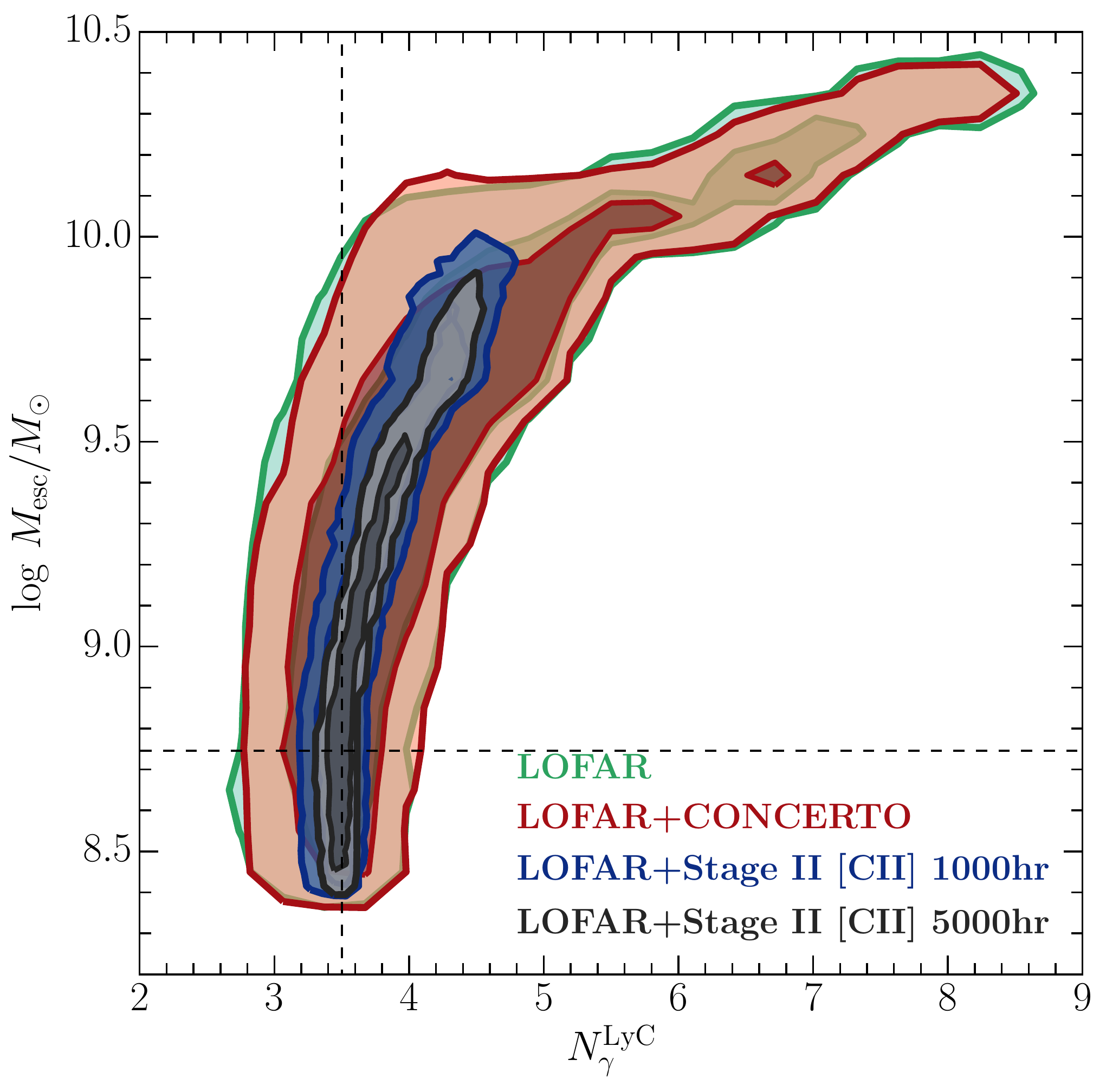} &
    \includegraphics[width=0.45\textwidth]{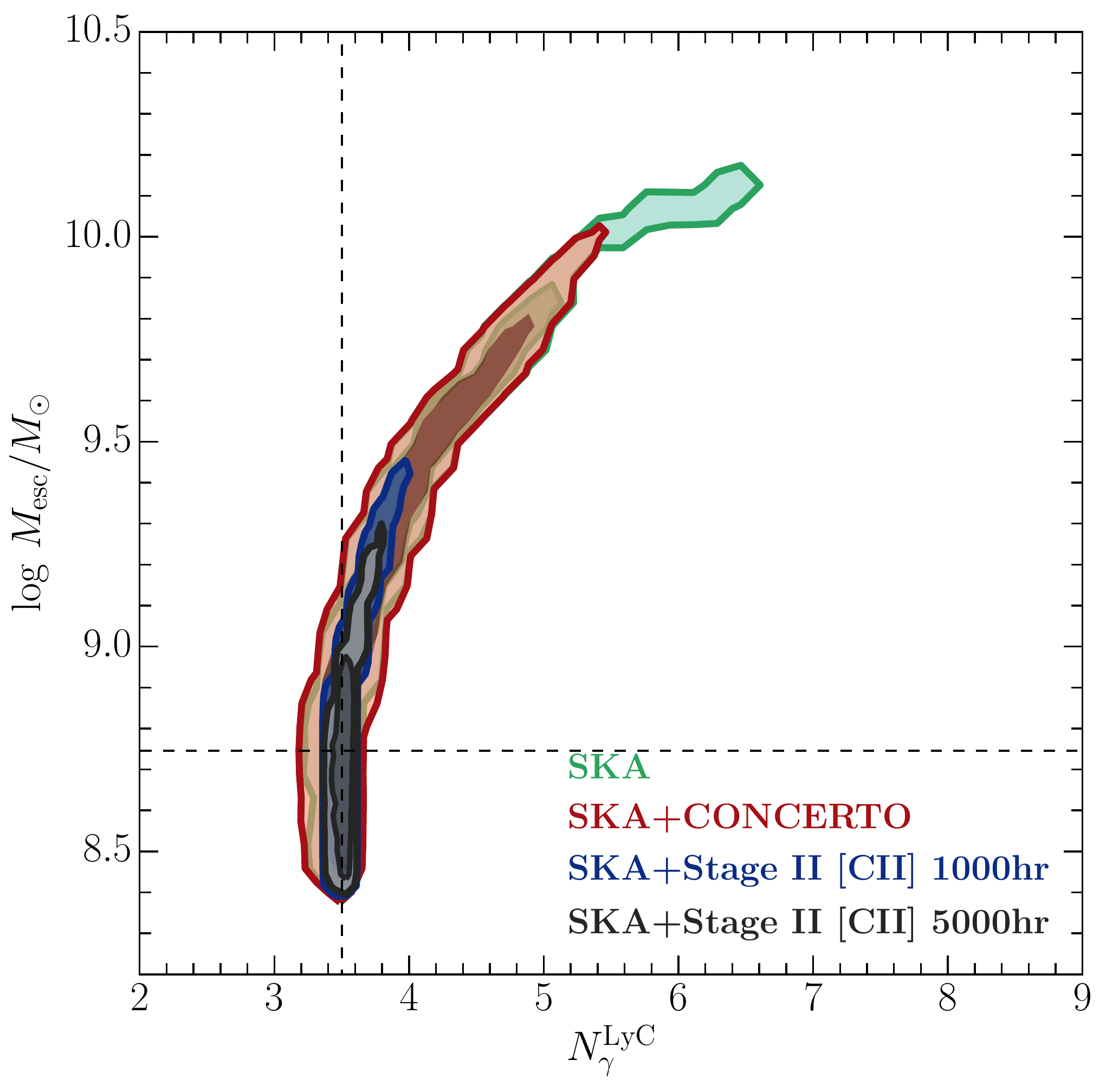} \\
    \includegraphics[width=0.45\textwidth]{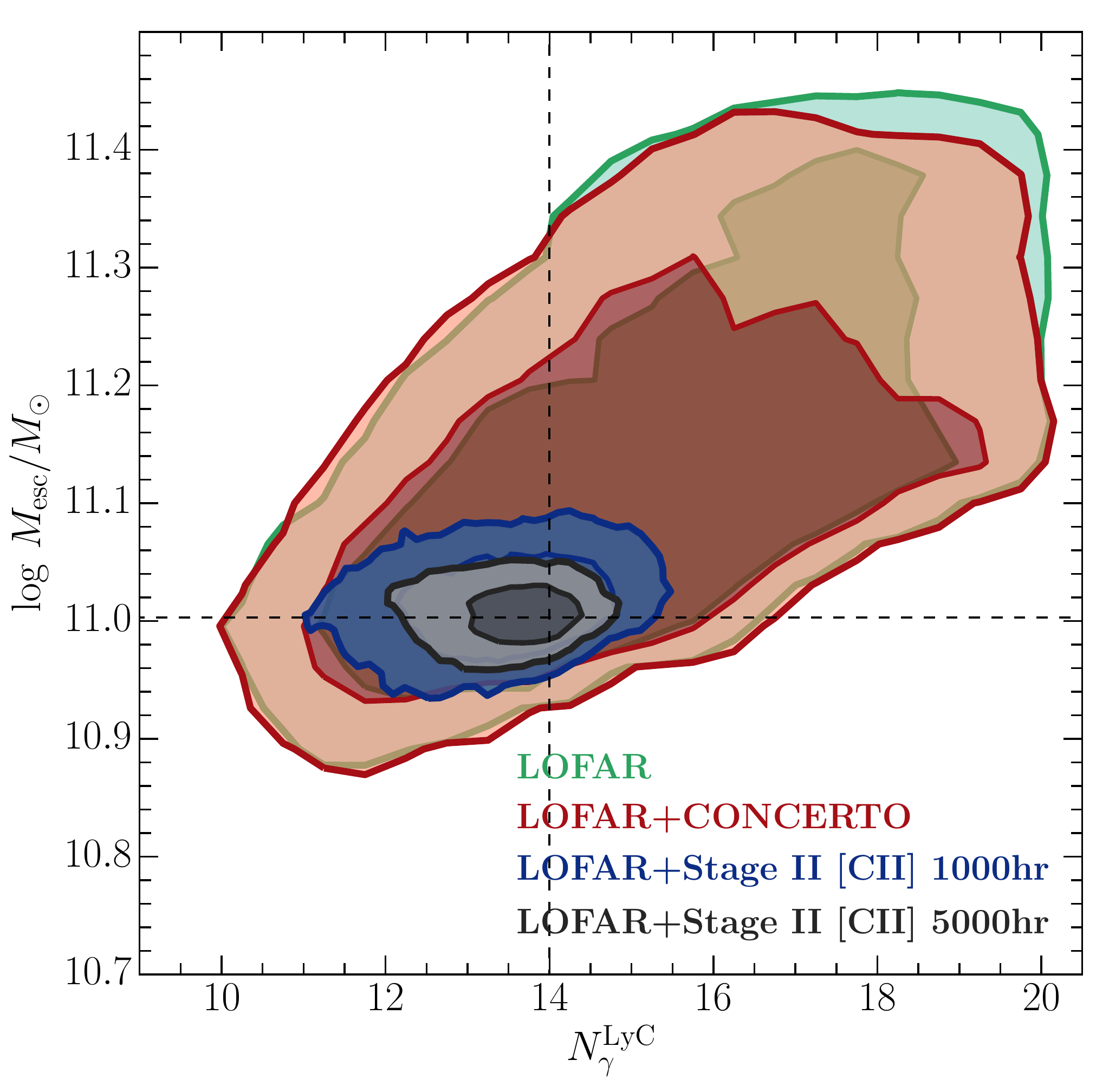} &
    \includegraphics[width=0.45\textwidth]{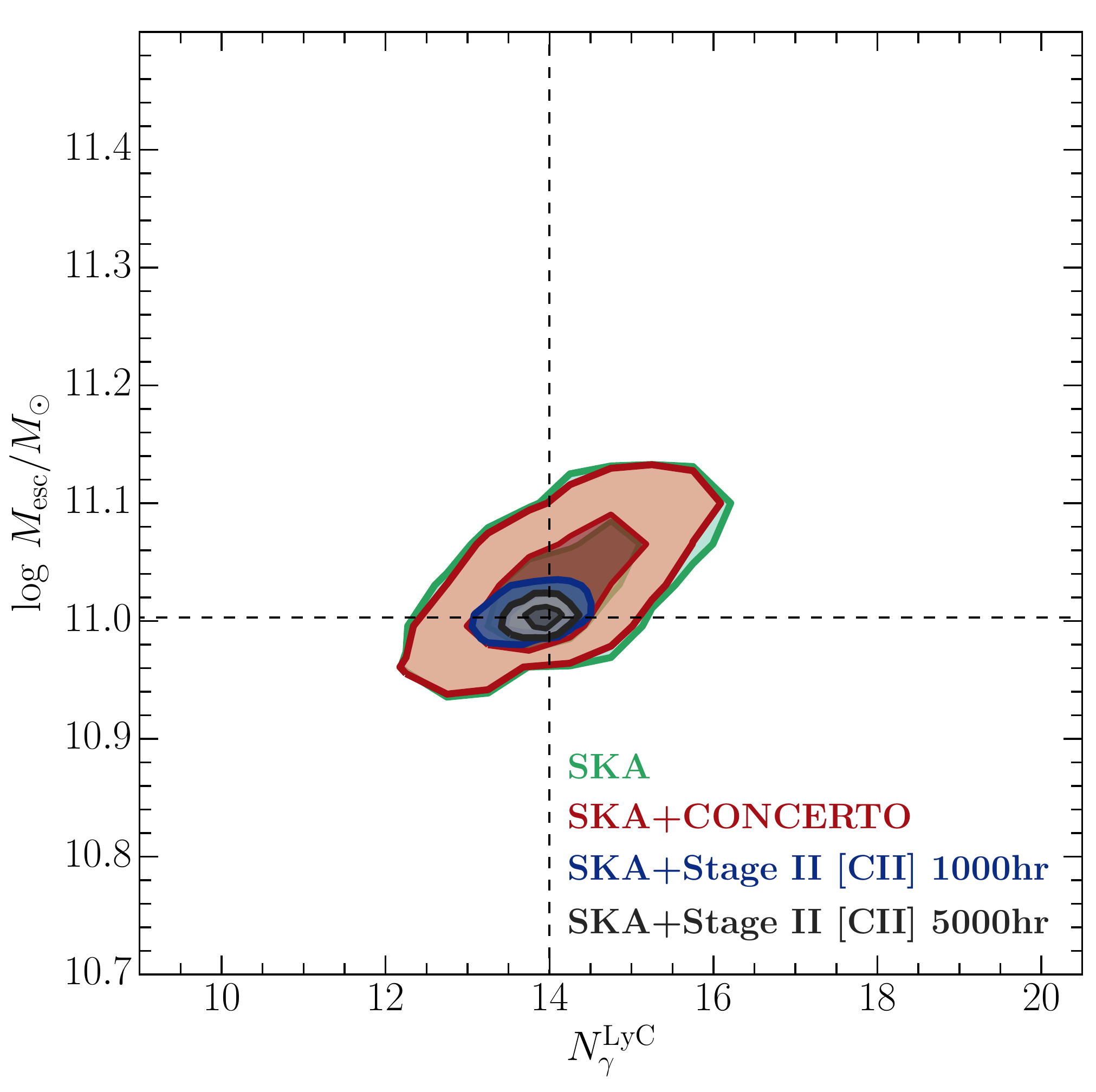} \\
  \end{tabular}
  \caption{Constraints on astrophysical parameters from mock
    measurements of the 21~cm power spectrum and the \c2-21cm cross
    power spectrum.  Panels in the top row describe the
    reionization scenario with low-mass sources, and those in the bottom
    row describe the reionization scenario with high-mass sources.  The two
    panels in each case refer to the use of LOFAR and SKA for the
    21~cm power spectrum measurement.  The dashed lines show the
    location of the ``true'' values of the parameters.  The green
    contours show the 1$\sigma$ and 2$\sigma$ constraints when only
    21\,cm power spectrum data is used.  Contours in other colours
    show constraints obtained when the \c2-21cm cross power spectrum
    data is added to the analysis. }
  \label{fig:forecasts}
\end{figure*}

\subsection{CII-21cm cross power spectrum sensitivity}
We calculate the uncertainty on the cross power spectrum of \c2 with 21~cm following \cite{2012ApJ...745...49G},
\begin{equation}
  \mathrm{var}[P_\mathrm{CII,21}(k,z)]=\frac{1}{2}\left[\frac{P^2_{21,\mathrm{CII}}+P^\mathrm{total}_{21}(k,z)P^\mathrm{total}_\mathrm{CII}}{N_\mathrm{m}(k,z)}\right],  
\end{equation}
where
\begin{equation}
  P^\mathrm{total}_{21}(k,z) = P_\mathrm{21}(k,z)+P_\mathrm{21}^\mathrm{N}(k,z),
\end{equation}
and
\begin{equation}
    P^\mathrm{total}_\mathrm{CII}(k,z) = P_\mathrm{CII}(k,z)+P_\mathrm{CII}^\mathrm{N}(k,z).
\end{equation}

The right-hand-side columns in Figures~\ref{fig:ps_gal} and
\ref{fig:ps_agn} show the errors on the cross power spectra for
CONCERTO-LOFAR (cyan) and CONCERTO-SKA (blue) combinations.  In both
cases, a high signal-to-noise detection of the cross power spectrum is
unlikely at least for scales smaller than $k\sim 0.1$~cMpc$^{-1}h$ at
$z=6$--$9$.  In the reionization model with low-mass sources, the
transition scale at which the cross power spectrum changes sign is at
$k\sim 5$~cMpc$^{-1}h$, which is out of the experimental reach.
However, as discussed in the previous section, the transition scale is
much larger, $k\sim 0.3$~cMpc$^{-1}h$, in the case of reionization by
high-mass sources.  This allows a detection of this scale, at least at
redshifts $z=6$ and $7$.  Figure~\ref{fig:ps_gal_ccat} shows errors on
the cross power spectra for LOFAR (cyan) and SKA (blue) combined with
our Stage~II
\c2\ experiment.  As expected the sensivities are enhanced now to
scales $k\sim 5$~cMpc$^{-1}h$ for LOFAR and $k>6$~cMpc$^{-1}h$ for SKA
at $z=7$.  Note that as the 21~cm signal originates in the extended
IGM, the shot noise contribution to the 21~cm power spectrum and the
\c2-21cm cross power spectrum is subdominant
\citep{2016MNRAS.463.2583K} and is not computed here.

\section{Forecasts for constraints}
\label{sec:forecasts}
We now consider the constraints that can be obtained for astrophysical
parameters related to reionization from measurements of (\textit{a})
the 21~cm power spectrum alone, and (\textit{b}) the 21~cm power
spectrum and the \c2-21cm cross power spectrum.  A variety of
astrophysical parameters determine the \c2 and 21\,cm emission from
the high-redshift universe.  As such \c2 and 21\,cm experiments can
potentially constrain all of these. However, for simplicity, we
consider only two parameters.  We consider a scenario in which haloes
down to the mass corresponding to the atomic hydrogen cooling limit
$T_\mathrm{vir}=10^4$~K produce \c2 emission, but only haloes with
mass $M>M_\mathrm{esc}$ have a non-zero Lyman-continuum photon escape
fraction.  Our simulation resolves haloes close to the atomic hydrogen
cooling limit.  Thus, this scenario assumes that all haloes in our
simulation are able to produce \c2 emission, but only massive haloes
with mass $M>M_\mathrm{esc}$ participate in reionization of the IGM.
The second parameter of our model is $N_\gamma^\mathrm{LyC}$, which
appears in Equation~(\ref{eqn:ngammam}) and sets number $N_\gamma$ of
ionizing photons produced by a halo.  Our two parameters,
$M_\mathrm{esc}$ and $N_\gamma^\mathrm{LyC}$ thus set the minimum mass
of haloes that produce ionizing photons and their Lyman-continuum
brightness, respectively.  The dependence of the Lyman-continuum
photon escape fraction on the halo mass is not well-understood. Our
choice of these parameters is therefore a simple proof of concept.
Nonetheless, some simple radiative transfer models in the literature
do suggest that Lyman-continuum photons are able to escape from a
narrow range of halo masses \citep{2013MNRAS.431.2826F}.  Our
parameterisation describes this possibility.

To assess the capability of observations to constrain the parameters
$M_\mathrm{esc}$ and $N_\gamma^\mathrm{LyC}$, we create mock power
spectra with experimental uncertainties and derive posterior
probability distributions for these parameters using MCMC.  This
approach is similar to that considered, for instance, for 21\,cm
experiments by \citet{2015MNRAS.449.4246G}.  We consider two mock
observations of the 21\,cm power spectrum and the \c2-21cm cross power
spectrum.  In one of these mocks, our two parameters have values
$M_\mathrm{esc}=5.56\times 10^8$~M$_\odot$ and
$N_\gamma^\mathrm{LyC}=3.5$.  This corresponds to the power spectra
shown in Figure~\ref{fig:ps_gal}.  The associated uncertainties are
also those shown in Figure~\ref{fig:ps_gal}.  For the second mock
observation, the mock measurements and associated errors are the power
spectra shown in Figure~\ref{fig:ps_agn}.  This mock data has
$M_\mathrm{esc}=10^{11}$~M$_\odot$ and $N_\gamma^\mathrm{LyC}=14$.  As
the shot noise dominates the total power, we only fit the clustering
power spectrum in this exercise.

For each of the mock datasets, we infer the posterior distributions
for $M_\mathrm{esc}$ and $N_\gamma^\mathrm{LyC}$ by writing a Gaussian
likelihood for the data as
\begin{multline}
  \log\mathcal{L}(\Delta^2|M_\mathrm{esc},N_\gamma^\mathrm{LyC}) \propto
  -\frac{1}{2}\sum_i\log\left(2\pi\sigma^2(k_i)\right) \\- \sum_i
  \frac{(\Delta^2_\mathrm{mock}(k_i)-\Delta^2_\mathrm{model}(k_i,M_\mathrm{esc},N_\gamma^\mathrm{LyC}))^2}{2\sigma^2(k_i)},   
\end{multline}
where $\Delta^2$ denotes the power spectrum or the cross-power
spectrum, as the case may be, the index $i$ runs over the $k$-bins,
and $\sigma$ is the error on the mock observation at wavenumber $k_i$,
estimated for various experiments following the procedure described in
Sections~\ref{sec:c2sens} and \ref{sec:21cmsens}.  We then explore the
capability of our model to identify the parameters used to create the
mock data, by inferring the values of these parameters in a Bayesian
fashion.  We use Markov Chain Monte Carlo to sample the posterior
distributions of the parameters, using a modified version of the 21~cm
inference code 21CMMC \citep{2015MNRAS.449.4246G} to derive
distributions for the parameters $M_\mathrm{esc}$ and
$N_\mathrm{min}^\mathrm{LyC}$ assuming wide, uniform priors.  For
given values of the parameters, we compute $\Delta^2_\mathrm{model}$
by first running our simulation (as described in
Section~\ref{sec:model} and Section~\ref{sec:cross-corr}; with the
power spectrum and cross power spectrum as defined in
Equations~\ref{eqn:ps} and \ref{eqn:crossps}) over a grid of points
in the parameter space and then linearly interpolating between the
values of the likelihood to get it at an arbitrary parameter value.
Our grid of models has 399 simulations.  It spans 19 values of
$N_\mathrm{min}^\mathrm{LyC}$ and 21 values of $M_\mathrm{esc}$.

The resultant posterior joint probability distributions are shown in
Figure~\ref{fig:forecasts}.  Panels in the top row describe the
low-mass reionization scenario, and those in the bottom row
describe the high-mass reionization scenario.  The two panels
in each case refer to the use of LOFAR and SKA for the 21\,cm power
spectrum measurement.  The dashed lines show the location of the
``true'' values of the parameters, which were used to produce the mock
data.  The green contours show the 1$\sigma$ and 2$\sigma$
constraints when only 21\,cm power spectrum data is used.  In this
case, there is a strong degeneracy in the two parameters in the
low-mass reionization scenario.  This degeneracy persists in
the high-mass case, although its magnitude is considerably
reduced.  Constraints in the high-mass case are good even with
the 21\,cm data alone, as the power spectrum has an enhanced amplitude
in this case, which allows for a high signal-to-noise measurement.
Contours in other colours in Figure~\ref{fig:forecasts} show
constraints obtained when the \c2-21cm cross power spectrum data is
added to the analysis.  We find that this considerably improves the
constraints for the Stage~II \c2\ experiment. With data from 1000~hr
and 5000~hr of the Stage~II experiment, the improvement in 1$\sigma$
constraints on $M_\mathrm{esc}$ relative to 21~cm measurements is by
factors of 3 and 10 respectively.  The improvement is of a comparable
magnitude in the high-mass case.  The constraints also show a
modest improvement when SKA measurements are considered instead of
LOFAR.  Due to low signal-to-noise, 1500~hr data from CONCERTO do not
result in a significant improvement in the constraints.

\section{Conclusions}
\label{sec:conclusions}

We have outlined the prospects of intensity mapping the epoch of
reionization using the redshifted 21\,cm line and the \c2 emission
line from high-redshift galaxies. We have modelled the galaxy line
emissions using a semi-analytical model.  Using a high dynamic range
cosmological simulation, we found that on large scales of $\gtrsim
60$~cMpc$/h$ at redshift $z=6$ the spherically averaged power spectrum
of the \c2 line emission have values of $\Delta^2\sim
10^5~(\mathrm{Jy}/\mathrm{sr})^2$ at $k\sim 0.2$~\kunit.  This value
reduces to about $10^3~(\mathrm{Jy}/\mathrm{sr})^2$ at $z\sim 9$.

We find that the \c2 power spectrum predicted in our model should be
detectable with the CONCERTO experiment up to $z\sim 8$ with a
signal-to-noise ratio of $\gtrsim 1$ at $k= 0.2$~\kunit.  A Stage-II
experiment with five times better sensitivity than CONCERTO should be
able to detect the \c2 power spectrum at even higher redshifts.  The
cross power spectrum of the \c2 and coeval 21\,cm signal from the
epoch of reionization would be valuable in many ways.  The scale at
which this cross power spectrum changes sign can contain the average
size of ionized regions, at least when the sources of reionization
coincide with the galaxies that produce the \c2 signal.  A detection
of this cross power spectrum could help in the removal of low-redshift
foregrounds from the 21~cm data.  The cross power spectrum will also
provide constraints on important astrophysical parameters.  We have
investigated the capability by analysing mock 21~cm power spectrum
data and \c2-21cm cross power spectrum data in a Bayesian way to
derive constraints under various experimental assumptions.  We find
that \c2-21cm correlation measurements can improve constraints on the
mass of reionization sources by factors of 3--10 beyond constraints
from 21~cm experiments alone.

\section*{Acknowledgements}

We acknowledge helpful discussions with Dongwoo Chung, Andrea
Pallottini and Ewald Puchwein, and also thank the anonymous referee
for their comments.  Support by ERC Advanced Grant 320596 `The
Emergence of Structure During the Epoch of Reionization' is gratefully
acknowledged.  We acknowledge PRACE for awarding us access to the
Curie supercomputer, based in France at the Tr\'es Grand Centre de
Calcul (TGCC). GL acknowledges financial support from the ``Programme
National de Cosmologie and Galaxies'' (PNCG) funded by
CNRS/INSU-IN2P3-INP, CEA and CNES, France.  This work used the DiRAC
Data Centric system at Durham University, operated by the Institute
for Computational Cosmology on behalf of the STFC DiRAC HPC Facility
(www.dirac.ac.uk). This equipment was funded by BIS National
E-infrastructure capital grant ST/K00042X/1, STFC capital grants
ST/H008519/1 and ST/K00087X/1, STFC DiRAC Operations grant
ST/K003267/1 and Durham University. DiRAC is part of the National
E-Infrastructure.

\bibliographystyle{mnras}
\bibliography{refs}

\bsp
\label{lastpage}
\end{document}